\documentclass{article} 
\usepackage{geometry}
\usepackage{multirow}
\usepackage{epsfig}
\usepackage{color} 
\usepackage{enumerate} 
\usepackage{graphicx}
\usepackage{graphics}
\usepackage{amsmath}
\usepackage{amsthm}
\usepackage{amssymb}
\usepackage{bm}
\usepackage{mdsymbol}
\usepackage{times} 
\usepackage{xspace}
\usepackage{hyperref}
\usepackage[square, sort, numbers]{natbib}
\usepackage{arydshln}
\usepackage{mathtools}

\newtheorem{theorem}{Theorem}[section]

\newtheorem{remark}[theorem]{Remark}

\newtheorem{assumption}{Assumption}

\newcommand{\R}{{\mathbb R}}

\newcommand{\finrema}{\hfill $\square$}

\begin{document}
\title{Novel Modelling and Control Strategies for a Steam Boiler under
  Fast Load Dynamics%
  \thanks{This research was supported partially 
     by the Australian Government through the Australian
    Research Council's Linkage Projects funding scheme (LP170100576)}}
\author{Diego S. Carrasco, Graham C. Goodwin, and Robert D. Peirce}
\maketitle

\begin{abstract}
  This paper describes a new nonlinear dynamic model for a natural
  circulation boiler. The model is based on physical principles,
  i.e. mass, energy and momentum balances. A systematic approach is
  followed leading to new insights into the physics of drum water
  level and downcomer mass flow. The model captures fast dynamic
  responses that are necessary to describe the operation of a boiler
  under highly variable load conditions. New features of the model
  include (i) a multi-compartment model for the risers, (ii) a new
  model for drum water level, and (iii) a new dynamic model for the
  flow of water in the downcomers. Implications of the model for
  control system design are explored in detail. Finally, the suggested
  improvements are validated in a sugar mill boiler.
\end{abstract}

\section{Introduction}
\label{sec:intro}
The current paper describes the outcomes of a three year project
carried out in collaboration with Wilmar Sugar at their Proserpine
Mill in Queensland, Australia. Sugar mills burn sugar cane residue
(bagasse) to produce steam in a boiler. The generated steam is
used for many purposes, including to power the factory, to
crystallise sugar, and to co-generate electricity.

Unlike boilers used in conventional gas or coal-fired power stations,
boilers in sugar mills are subject to large and rapidly changing
loads, e.g. when cane crushers are started or stopped. In addition,
the fuel (bagasse) has a highly variable calorific value due to the
different moisture content in the original cane. The result of these
two factors is that (i) boilers in sugar mills must necessarily cope
with large and rapid load changes,e.g. $50\%$ load change over the
span of~$6-8$ seconds, and (ii) the high moisture content in the
bagasse can make it very difficult to burn, thus impacting furnace
dynamics. As a consequence, boilers in sugar mills can experience
severe operational difficulties, including frequent stoppages due to
large drum water level excursions. To address the aforementioned
problems, a novel boiler model was developed aimed specifically at
capturing fast load dynamics. The model was then used to redesign
the associated control system.

Dynamic models for Boilers have appeared in the literature for many
years~\cite{anderson69,schulz73,dukelow86,kwatny96,Maffezzoni1997,majanne17}.
Early work focused on obtaining empirical models capable of describing
the internal dynamics with limited
accuracy~\cite{astrom72,Astrom1975}. Throughout the years, the focus
has shifted to develop models based on physical principles,
i.e. first-principle
models~\cite{bell+ast96,Maffezzoni1997,Astrom2000}. Models for boilers
have taken many different formats,
e.g. linear/nonlinear~\cite{Astrom2000,Tan2008}, high/low
order~\cite{R.D.Bell1987}, one/two fluid~\cite{Alobaid2016},
lumped/distributed parameter~\cite{Sunil2017,Astrom2000}, and have
also been tailored to different thermal power plant
technologies~\cite{Alobaid2016}.

In the seminal work of~\cite{Astrom2000}, a model using mass and energy
balances is described. Several simplifications were used when
developing this model, namely (i) a steady state equation for the
downcomer mass flow, and (ii) an assumption that steam quality varies
linearly as a function of height in the risers.  The more recent work
described in \cite{Sedic2014} derives a first principles model using
mass, energy and momentum conservation equations. However, other
simplifications are used when developing this model, including the
fact that pressure and other internal variables are reconstructed by
first order filters. It is shown in the current paper that the
simplifications and assumptions used in~\cite{Astrom2000,Sedic2014}
are not valid under rapidly changing load conditions.


Based on the above background, this paper presents a new nonlinear
model for natural a circulation boiler based entirely on physical
principles, including mass and energy balances and implications of
constant volume of the different sections of the boiler. The goal is
to develop a model that (i) captures the internal fast dynamics needed
to account for large and rapid load variations and fuel variability,
and (ii) is simple enough to support controller design. Another aspect
of the model development is that all assumptions are clearly and
explicitly stated. Thus they can be readily assessed for their
validity in specific cases.

The key new features of the model described in this paper are:
\begin{itemize}
\item A multi-compartment model is developed for the risers. It will be
  shown that the spatially distributed nature of boiling water in the
  risers plays a central role in drum water level dynamics under
  fast and large load changes.
\item A new model for drum water level is developed. The model gives
  rise to a key controller design insight, i.e. drum
  water level deviations are proportional to steam flow out of the boiler.
\item A new model for downcomer water flow in a natural circulation
    boiler is described. It will be shown that a general momentum
    balance approach provides a link between water flow and pressure
    derivatives.
\end{itemize}

The work presented here embellishes and extends the work
in~\cite{Astrom2000}. The pressure model turns out to be equivalent to
that presented in~\cite{Astrom2000}. The novelty of the current paper
lies in (i) the three key features described above, and (ii) the
manner in which the equations are used to describe the model. It will
be seen that, through simple algebraic manipulations, the final
model format allows for easy understanding and
facilitates the design and (re)-tuning of controllers.

\begin{figure}[h]
\centering
\resizebox{0.5\columnwidth}{!}{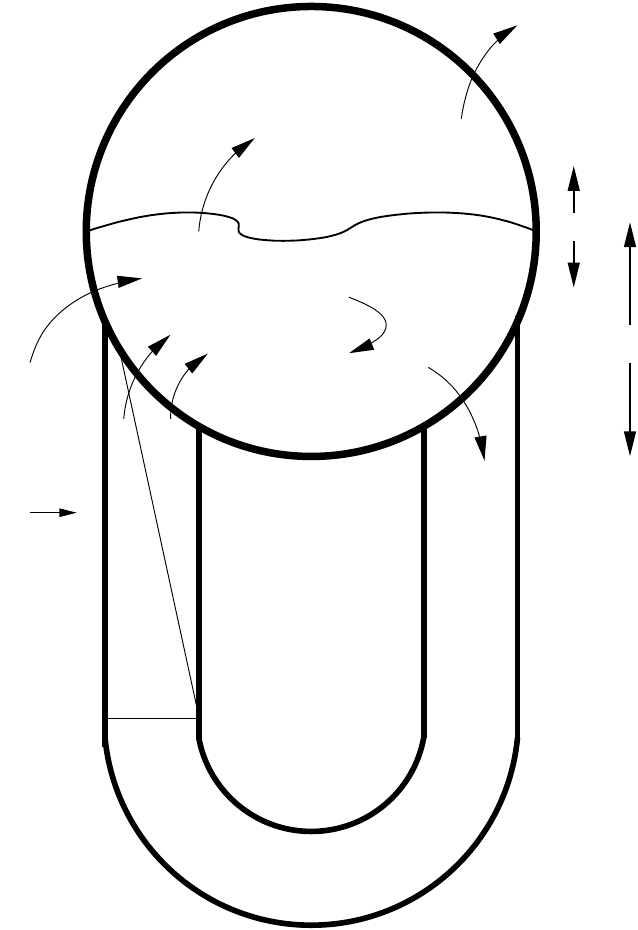}
       \caption{Cross section of boiler}
       \label{fig:boiler}
     \end{figure}

The remainder of the paper is organised as follows: In
Section~\ref{sec:drum} the equations for mass balance, energy balance
and volume constraints on the drum are described.  In
Section~\ref{sec:risers}, the equations for mass balance, energy
balance and volume constraints on the risers are described.  In
Section~\ref{sec:pmodel}, a model for boiler pressure is developed.
In Section~\ref{sec:dwlmodel}, a model for drum water level is
developed. It is shown that further equations are necessary.  In
Section~\ref{sec:newrisers}, the equations corresponding to spatial
discretisation of the risers are developed. An additional assumption
of homogeneous mixing in the risers is introduced.  In
Section~\ref{sec:momentum}, an equation for the mass flow of water in
the downcomers is developed based on momentum balance.  In
Section~\ref{sec:superheater}, a model for a superheater based on
constant volume, plus energy and mass balances is developed.  In
Section~\ref{sec:simulations}, key consequences of the new model are
discussed and simulations are presented highlighting the key new
features. In Section~\ref{sec:implications}, the implications of the
new model relative to the control of drum water level are explored. In
Section~\ref{sec:exp}, experimental results obtained from a boiler
operating in the sugar industry are presented. This results confirm
the benefits of the new model and the associated control strategies. In
Section~\ref{sec:conclusions}, conclusions are drawn.
For ease of reference, a list of the variables used throughout the paper
is presented in Table~\ref{tab:var}, and a schematic of the boiler is
given in Fig.~\ref{fig:boiler}.
\begin{table}[h]\label{tab:var}
  \centering
  \begin{tabular}[H]{cll}
\hline
Symbol &Description& See Eq.\\
\hline
$\alpha^i$&Steam quality in riser section~$i$&\eqref{eq:alphai}\\
$\delta$&Drum water level deviations&\eqref{eq:delta}\\
$f_\ell$&Mass flow of water converted into steam&\eqref{eq:fell}\\
$f_s$&Steam mass flow from risers into drum&\eqref{eq:fsifinal} and~\eqref{eq:fs}\\
$f_w$&Water mass flow from risers into drum&\eqref{eq:fwifinal} and~\eqref{eq:fw}\\
$f_{w^*}$& Water mass flow from drum into risers& \eqref{eq:fw*}\\
    $h_s$&Enthalpy of steam&Steam tables\\
    $h_w$&Enthalpy of water&Steam tables\\
    $h_s$&Enthalpy of feedwater&Assumed known\\
$M_s^D$&Mass of steam in the drum (above water line)&\eqref{eq:MsD}\\
$M_w^D$& Mass of water in the drum&\eqref{eq:MwD}\\
$M_s^R$& Mass of steam in the risers&\eqref{eq:MsR}\\
$M_w^R$& Mass of water in the risers&\eqref{eq:MwR}\\
$M_s^{BW}$&Mass of steam below the water line&\eqref{eq:MsBW}\\
$P$&Drum Pressure&\eqref{eq:Pdot}\\
$Q^B$&Heat flow used to turn water into steam&Control variable\\
$q_f$&Mass flow of feedwater&Control variable\\
$q_s$&Mass flow of steam exiting the drum&Assumed known\\
$\rho_s$&Steam density&Steam tables\\
$\rho_w$&Water density&Steam tables\\
$V^D$&Total volume of the drum&Assumed known\\
$V_s^D$&Volume of steam in the drum&\eqref{eq:VD}\\
$V_w^D$&Volume of water in the drum&\eqref{eq:VD}\\
$V_s^{BW}$&Volume of steam below the water line&\eqref{eq:DWL1}\\
$V^R$&Total volume of the risers&Assumed known\\
$V_s^R$&Volume of steam in the risers&\eqref{eq:VR}\\
$V_w^R$&Volume of water in the risers&\eqref{eq:VR}\\
\hline
  \end{tabular}
  \caption{List of Variables}
  \label{tab:ref}
\end{table}

\newpage
\section{Steam Drum}\label{sec:drum}
In this Section, mass balance, energy balance and volume constraint
equations for the drum section of the boiler will be
derived. Algebraic manipulations of the latter two will give rise to
equations used in the final model. Two assumptions are needed for the
subsequent derivations, namely:

\begin{assumption}\label{ass:water_sat}
Water in the boiler is at its saturation temperature~$T_s$.\finrema
\end{assumption}
\begin{assumption}\label{ass:metal_sat}
  The temperature of the metal,~$T_m$, is the same as
  the saturation temperature,~$T_s$.\finrema
\end{assumption}

\subsection{Mass Balance}
The mass of steam and water in the drum satisfy conservation
equations. In particular, the time rate of change of mass contained in
an open system is equal to the difference between mass inflow and
outflow of the system.  This leads to:
\begin{align}
  \dot{M}_s^D ={}& f_s-q_s-f_{cd}\label{eq:MsD}\\
  \dot{M}_w^D ={}& q_f+f_w-f_w^*+f_{cd}\label{eq:MwD}
\end{align}
where~${M}_s^D,~{M}_w^D$ denote the mass of steam in the drum, and the
mass of water in the drum, respectively, and
where~$f_s,~q_s,~q_f,~f_w,~f_w^*,f_{cd}$ denote the mass flow of steam
from the risers into the drum, the mass flow of steam out of the drum,
the mass flow of feedwater into the drum, the mass flow of water from
the top of the risers into the drum, the mass flow of water from the
drum into the downcomers, and the mass flow due to steam condensation,
respectively.

\begin{remark}
  Note that $M_s^D$ is defined as the total mass of steam in the
  drum, i.e., it includes both the mass of steam above and below
  water.\finrema
\end{remark}

\subsection{Energy Balance}
The time rate of change of the energy contained in
an open system is equal to the difference between energy inflows and
outflows of the system. The energy contained in the drum is given by
the energy contained in the masses of water and steam in the drum,
plus the energy contained in the metal walls of the drum. The
following energy balance equation results from these considerations:
\begin{align}
  \dfrac{d}{dt} \left\{     M_s^Du_s + M_w^Du_w +  M_m^DC_pT_m \right\}
  ={}& (f_s-q_s)h_s + (f_w-f_w^*)h_w + q_fh_w^f \label{eq:drumEB}
\end{align}
where~$u_s~,u_w,~M_m^D,~C_p,~T_m$ denote the internal energy of steam,
internal energy of water, mass of metal in the drum, heat capacity of
metal, and temperature of metal,
respectively. Also~$h_s,~h_w,~h_w^f$ denote the enthalpy of
steam, enthalpy of water, and enthalpy of feedwater, respectively.

By definition, internal energy is related to enthalpy and pressure
by~$u=h-P/\rho$. Substituting into the left hand side
of~\eqref{eq:drumEB} leads to:
\begin{align}
  \dfrac{d}{dt} \left\{ M_s^Dh_s + M_w^Dh_w -
  \left( \dfrac{M_s^D}{\rho_s} + \dfrac{M_w^D}{\rho_w} \right)P+
  M_m^DC_pT_m \right\} 
  ={}& (f_s-q_s)h_s + (f_w-f_w^*)h_w + q_fh_w^f \nonumber
\end{align}
Noting that
\begin{align}
  \dfrac{M_s^D}{\rho_s} + \dfrac{M_w^D}{\rho_w}
  ={}& V_s^D + V_w^D = V^D \nonumber
\end{align}
then,
\begin{align}
  \dfrac{d}{dt} \left\{ M_s^Dh_s + M_w^Dh_w - V^D P +  M_m^DC_pT_m \right\} 
  ={}& (f_s-q_s)h_s + (f_w-f_w^*)h_w + q_fh_w^f   \nonumber
\end{align}
Expanding the LHS, leads to:
\begin{align}
\dot{M}_s^Dh_s + M_s^D\dot{h}_s + \dot{M}_w^Dh_w +M_w^D\dot{h}_w - V^D
  \dot{P} +  M_m^DC_p\dot{T}_m
   ={}& (f_s-q_s)h_s + (f_w-f_w^*)h_w + q_fh_w^f \label{eq:DEB1}
\end{align}
Considering Assumption~\ref{ass:water_sat}, it follows that:
\begin{align}\label{eq:hdot}
  \dot{h} ={}& \dfrac{dh}{dt} =
  \dfrac{\partial{}h}{\partial{}P}\cdot \dfrac{dP}{dt} =
  \dfrac{\partial{}h}{\partial{}P}\cdot \dot{P}
\end{align}
Considering Assumption~\ref{ass:metal_sat}, it follows that:
\begin{align}
  \label{eq:tdot}
    \dot{T}_m \approx{}& \dfrac{dT_s}{dt} =
  \dfrac{\partial{}T_s}{\partial{}P}\cdot \dfrac{dP}{dt} =
  \dfrac{\partial{}T_s}{\partial{}P}\cdot \dot{P}
\end{align}
Substituting~\eqref{eq:hdot} and~\eqref{eq:tdot} into~\eqref{eq:DEB1},
and using the mass balance equations~\eqref{eq:MsD}
and~\eqref{eq:MwD}, leads to:
\begin{align}
  (f_s-q_s-f_{cd})h_s + (q_f+f_w-f_w^*+f_{cd})h_w 
  +&\left(  M_s^D\dfrac{\partial{}h_s}{\partial{}P} +  M_w^D\dfrac{\partial{}h_w}{\partial{}P} - V^D
 +M_m^DC_p \dfrac{\partial{}T_s}{\partial{}P} \right) \dot{P}  \nonumber\\
   &={} (f_s-q_s)h_s + (f_w-f_w^*)h_w + q_fh_w^f\nonumber
\end{align}
Cancelling the common terms on both sides, and introducing the
following definition
\begin{align}
  \label{eq:K1}
  K_1 \triangleq{}&  M_s^D\dfrac{\partial{}h_s}{\partial{}P} +
                    M_w^D\dfrac{\partial{}h_w}{\partial{}P} - V^D
                    +M_m^DC_p \dfrac{\partial{}T_s}{\partial{}P},
\end{align}
leads to:
\begin{align}
  (-f_{cd})h_s+(q_f+f_{cd})h_w 
  +K_1 \dot{P} 
   &={}  q_fh_w^f\nonumber
\end{align}
Rearranging the above equation leads to:
\begin{align}
  \label{eq:fcd}
 f_{cd} ={}& (h_s-h_w)^{-1} \left( K_1\dot{P} + q_f ( h_w-h_w^f)\right)  
\end{align}
This equation will form part of the final model 
described in Section~\ref{sec:pmodel}.

\subsection{Volume Constraint}
The total volume of the steam drum is
constant. This leads to the following equation:
\begin{align}
  V^D ={}& V_s^D + V_w^D\label{eq:VD}
\end{align}
where~$V^D, V_s^D,~V_w^D$ denote the volume of the drum, the volume of
steam in the drum (both above and below the water line), and the
volume of water in the drum, respectively.

Since the volume of the drum is constant, using~\eqref{eq:VD} and
taking its derivative leads to:
\begin{align}
  0 ={}& \dfrac{d}{dt} \left\{ V_s^D+V_w^D \right\} \nonumber\\
   ={}& \dfrac{d}{dt} \left\{ \dfrac{M_s^D}{\rho_s} + \dfrac{M_w^D}{\rho_w} \right\} \nonumber
\end{align}
where the fact that~$V=M/\rho$ has been used, where~$M,~\rho$ denote
mass and density, respectively. Expanding the derivative leads to:
\begin{align}
0  ={}& \dfrac{\dot{M}_s^D}{\rho_s} - \dfrac{M_s^D}{\rho_s^2}\cdot{}\dot{\rho}_s
        + \dfrac{\dot{M}_w^D}{\rho_w} -
        \dfrac{M_w^D}{\rho_w^2}\cdot{}\dot{\rho}_w\label{eq:maniVD1} 
\end{align}
A consequence of Assumption~\ref{ass:water_sat} is that density is a
function of pressure only. Thus,
\begin{align}\label{eq:rhodot}
  \dot{\rho} ={}& \dfrac{d\rho}{dt} =
  \dfrac{\partial\rho}{\partial{}P}\cdot \dfrac{dP}{dt} =
  \dfrac{\partial\rho}{\partial{}P}\cdot \dot{P}
\end{align}
where~$P$ denotes the pressure of the boiler. Using~\eqref{eq:rhodot},
and substituting the mass balance equations~\eqref{eq:MwD} and~\eqref{eq:MsD}
into~\eqref{eq:maniVD1} leads to:
\begin{align}
  0  ={}&  \dfrac{f_s-q_s-f_{cd}}{\rho_s} +\dfrac{ q_f+f_w-f_w^*+f_{cd}}{\rho_w} 
          - \left(
          \dfrac{M_s^D}{\rho_s^2}\dfrac{\partial\rho_s}{\partial{}P} +
          \dfrac{M_w^D}{\rho_w^2}\dfrac{\partial\rho_w}{\partial{}P} \right)\cdot \dot{P}\nonumber
\end{align}
Introducing the following definition
\begin{align}
  \label{eq:C1}
  C_1 \triangleq{}&
                    \dfrac{M_s^D}{\rho_s^2}\dfrac{\partial\rho_s}{\partial{}P}
                    + \dfrac{M_w^D}{\rho_w^2}\dfrac{\partial\rho_w}{\partial{}P}
\end{align}
and rearranging terms, leads to:

\begin{align}\label{eq:CVD}
  \dot{P}  ={}& C_1^{-1} \left( \dfrac{-q_s-f_{cd}}{\rho_s} +\dfrac{
                q_f-f_w^*+f_{cd}}{\rho_w} 
                + \dfrac{f_s}{\rho_s} + \dfrac{f_w}{\rho_w}\right)
\end{align}
This equation will also form part of the final model described in
Section~\ref{sec:pmodel}.

\newpage
\section{Risers}\label{sec:risers}
In a similar fashion to Section~\ref{sec:drum}, mass balance, energy
balance and volume constraint equations for the risers are presented
in this Section. Algebraic manipulations of the latter two will give
rise to equations used in the final model.

\subsection{Mass Balance}
Mass balance in the risers leads to:
\begin{align}
  \dot{M}_s^R ={}& f_\ell - f_s\label{eq:MsR}\\
  \dot{M}_w^R ={}& f_w^* - f_w - f_\ell\label{eq:MwR}
\end{align}
where~${M}_s^R,~{M}_w^R$ denote the mass of steam in the
risers and the mass of water in the risers, respectively, and
where~$f_s,~f_w,~f_w^*,~f_\ell$ denote the mass flow of
steam from the risers into the drum, the mass flow of water
from the top of the risers into the drum, the mass flow of water from the drum
into the downcomers, and the mass flow of water converted into steam
in the risers, respectively.

\begin{remark}
  The above equations consider the risers as one section. That is
  sufficient for the moment. However, in Section~\ref{sec:newrisers}, a
  multiple-compartment model for the risers will be introduced. This will
  be necessary to more accurately describe the drum water level
  dynamics.\finrema
\end{remark}

\subsection{Energy Balance}
Energy balance in the risers leads to:
\begin{align}
  \dfrac{d}{dt} \left\{     M_s^Ru_s + M_w^Ru_w +  M_m^RC_pT_m  \right\}
  ={}& - f_sh_s + (f_w^*-f_w)h_w +Q^B \label{eq:risersEB}
\end{align}
where~$u_s~,u_w,~M_m^R,~C_p,~T_m$ denote the internal energy of steam,
internal energy of water, mass of metal in the risers, heat capacity
of metal, and temperature of metal, respectively. Also~$h_s,~h_w,~Q^B$
denote the enthalpy of steam, enthalpy of water, and heat flow used
for boiling water in the risers, respectively.

\label{sec:derivREB}
Substituting~$u=h-P/\rho$ into the left hand
side of~\eqref{eq:risersEB} leads to:
\begin{align}
  \dfrac{d}{dt} \left\{ M_s^Rh_s + M_w^Rh_w -
  \left( \dfrac{M_s^R}{\rho_s} + \dfrac{M_w^R}{\rho_w} \right)P+
  M_m^RC_pT_m \right\} 
  ={}&- f_sh_s + (f_w^*-f_w)h_w +Q^B \nonumber
\end{align}
Again, noting that
\begin{align}
  \dfrac{M_s^R}{\rho_s} + \dfrac{M_w^R}{\rho_w}
  ={}& V_s^R + V_w^R = V^R \nonumber
\end{align}
Then,
\begin{align}
  \dfrac{d}{dt} \left\{ M_s^Rh_s + M_w^Rh_w - V^R P +  M_m^RC_pT_m \right\} 
  ={}& - f_sh_s + (f_w^*-f_w)h_w +Q^B   \nonumber
\end{align}
Expanding the LHS leads to
\begin{align}
\dot{M}_s^Rh_s + M_s^R\dot{h}_s + \dot{M}_w^Rh_w +M_w^R\dot{h}_w - V^R
  \dot{P} +  M_m^RC_p\dot{T}_m
   ={}& - f_sh_s + (f_w^*-f_w)h_w +Q^B \label{eq:REB1}
\end{align}
In view of Assumption~\ref{ass:water_sat}, equations~\eqref{eq:hdot}
and~\eqref{eq:tdot} can be used here. In addition, substituting the
mass balance equations,~\eqref{eq:MsR} and~\eqref{eq:MwR},
into~\eqref{eq:REB1}, leads to:
\begin{align}
  (f_\ell-f_s)h_s + (f_w^*-f_w-f_\ell)h_w 
  +&\left(  M_s^R\dfrac{\partial{}h_s}{\partial{}P} +
     M_w^R\dfrac{\partial{}h_w}{\partial{}P} - V^R 
 +M_m^RC_p \dfrac{\partial{}T_s}{\partial{}P} \right) \dot{P}  \nonumber\\
   &={} - f_sh_s + (f_w^*-f_w)h_w +Q^B\nonumber
\end{align}
Cancelling the common terms on both sides, and introducing the
following definition
\begin{align}
  \label{eq:K2}
  K_2 \triangleq{}&  M_s^R\dfrac{\partial{}h_s}{\partial{}P} +
                    M_w^R\dfrac{\partial{}h_w}{\partial{}P} - V^R 
 +M_m^RC_p \dfrac{\partial{}T_s}{\partial{}P},
\end{align}
leads to:
\begin{align}
  f_\ell h_s - f_\ell h_w +K_2 \dot{P} = Q^B\nonumber
\end{align}
Rearranging the above equation yields:
\begin{align}\label{eq:fell}
 f_\ell  ={}&\left(h_s - h_w\right)^{-1}  \left( - K_2\dot{P} + Q^B  \right)
\end{align}
This equation will form part of the model for boiler pressure
described in Section~\ref{sec:pmodel}.

\subsection{Volume Constraint}
The volume of the risers is constant, leading to the
following equation:
\begin{align}
  V^R ={}&  V_s^R + V_w^R\label{eq:VR}
\end{align}
where~$V^R,~V_s^R,~V_w^R$ denote the volume of the risers, the volume
of steam in the risers, and the volume of water in the risers, respectively.

\label{sec:derivCVR} 
In the same fashion as the volume constraint for the drum, taking the
derivative of~\eqref{eq:VR} leads to:
\begin{align}
  0 ={}& \dfrac{d}{dt} \left\{ V_s^R+V_w^R \right\} \nonumber\\
   ={}& \dfrac{d}{dt} \left\{ \dfrac{M_s^R}{\rho_s} +
        \dfrac{M_w^R}{\rho_w} \right\} \nonumber\\
  ={}& \dfrac{\dot{M}_s^R}{\rho_s} - \dfrac{M_s^R}{\rho_s^2}\cdot{}\dot{\rho}_s
        + \dfrac{\dot{M}_w^R}{\rho_w} -
       \dfrac{M_w^R}{\rho_w^2}\cdot{}\dot{\rho}_w\nonumber\\
     ={}& \dfrac{ f_\ell - f_s}{\rho_s} + \dfrac{f_w^* - f_w - f_\ell}{\rho_w} 
          - \left( \dfrac{M_s^R}{\rho_s^2}\dfrac{\partial\rho_s}{\partial{}P}
        + \dfrac{M_w^R}{\rho_w^2}\dfrac{\partial\rho_w}{\partial{}P}
          \right)\cdot \dot{P}\nonumber
\end{align}
Introducing the following definition
\begin{align}
  \label{eq:C2}
  C_2 \triangleq{}& \dfrac{M_s^R}{\rho_s^2}\dfrac{\partial\rho_s}{\partial{}P}
        + \dfrac{M_w^R}{\rho_w^2}\dfrac{\partial\rho_w}{\partial{}P}
\end{align}
and rearranging terms, leads to:
\begin{align}\label{eq:CVR}
  \dfrac{f_s}{\rho_s} + \dfrac{f_w}{\rho_w}
  ={}& \dfrac{ f_\ell}{\rho_s} + \dfrac{f_w^* - f_\ell}{\rho_w} 
          - C_2\cdot \dot{P}
\end{align}
This equation will also form part of the final model described in
Section~\ref{sec:pmodel}.

\section{A Model for Boiler Pressure}\label{sec:pmodel}
So far, the model for the boiler dynamics comprises: (i) the four
conservation of mass equations,~\eqref{eq:MsD}--\eqref{eq:MwD}
and~\eqref{eq:MsR}--\eqref{eq:MwR}, (ii) the two equations derived
from the constant volume of the drum and risers,~\eqref{eq:CVD}
and~\eqref{eq:CVR}, and (iii) the two equations derived from energy
balance in the drum and in the risers,~\eqref{eq:fcd}
and~\eqref{eq:fell}. Here an expression for boiler pressure,~$P$, will
be derived based only on the latter four equations.
Substituting~\eqref{eq:CVR} into~\eqref{eq:CVD} leads to:
\begin{align*}
  \dot{P}  ={}& C_1^{-1} \left( \dfrac{-q_s-f_{cd}}{\rho_s}
                + \dfrac{ q_f-f_w^*+f_{cd}}{\rho_w}
                + \dfrac{ f_\ell}{\rho_s} + \dfrac{f_w^* - f_\ell}{\rho_w}
                - C_2\cdot \dot{P}\right)
\end{align*}
Grouping common terms for~$\dot{P}$ yields:
\begin{align}
  \label{eq:Pmodel1}
  \left( C_1 + C_2\right) \dot{P}
  ={}&
       \dfrac{-q_s-f_{cd}}{\rho_s} + \dfrac{ q_f+f_{cd}}{\rho_w}
       + \left( \dfrac{1}{\rho_s} - \dfrac{1}{\rho_w} \right) f_\ell
\end{align}
Substituting with~\eqref{eq:fell}, it follows that:
\begin{align*}
  \left( C_1 + C_2\right) \dot{P}
  ={}&
       \dfrac{-q_s-f_{cd}}{\rho_s} + \dfrac{ q_f+f_{cd}}{\rho_w}
       + \left( \dfrac{1}{\rho_s} - \dfrac{1}{\rho_w} \right)
  \left(h_s - h_w\right)^{-1}  \left( - K_2\dot{P} + Q^B  \right)
\end{align*}
The following variable is then introduced to simplify the equations:
\begin{align}
  \label{eq:C3}
  C_3 \triangleq \left( \dfrac{1}{\rho_s} - \dfrac{1}{\rho_w} \right)
       \left(h_s - h_w\right)^{-1}
\end{align}
Rearranging the above equation leads to:
\begin{align*}
  \left( C_1 + C_2 + C_3K_2\right) \dot{P}
  ={}&
       \dfrac{-q_s}{\rho_s} + \dfrac{ q_f}{\rho_w}
       + C_3  Q^B
       - \left( \dfrac{1}{\rho_s} - \dfrac{1}{\rho_w} \right) f_{cd}
\end{align*}
Using~\eqref{eq:fcd}, it follows that
\begin{align*}
  \left( C_1 + C_2 + C_3K_2\right) \dot{P}
  ={}&
       \dfrac{-q_s}{\rho_s} + \dfrac{ q_f}{\rho_w}
       + C_3  Q^B
       - C_3 \left( K_1\dot{P} + q_f ( h_w-h_w^f)\right)  
\end{align*}
Finally, solving for~$\dot{P}$ yields:
\begin{align}\label{eq:Pdot}
    \dot{P}
  ={}&
\left( C_1 + C_2 + C_3 (K_1+ K_2) \right)^{-1}
       \left( \dfrac{ -q_s}{\rho_s} + \dfrac{q_f}{\rho_w}
       + C_3 \left(Q^B- q_f ( h_w-h_w^f) \right) \right)
\end{align}

In summary, equation~\eqref{eq:Pdot} is a differential equation
for~$P$, equations~\eqref{eq:fell} and~\eqref{eq:fcd} are algebraic
equations for~$f_\ell$ and~$f_{cd}$, respectively, and
equations ~\eqref{eq:MsD}--\eqref{eq:MwD}
and~\eqref{eq:MsR}--\eqref{eq:MwR} are differential equations
for the steam and water masses in the boiler. Together, they provide a
complete model for boiler pressure.

\begin{remark}
  Note that, to evaluate~\eqref{eq:Pdot}, it is not necessary to know
  the values of~$f_s,~f_w$ and~$f_w^*$. The variable~$C_3$ is determined
   only by the boiler pressure, while~$K_1+K_2$ and~$C_1+C_2$ depend
  on pressure and the \emph{total} mass of water and steam in the
  boiler, since~$\dot{M}_w^D+\dot{M}_w^R=q_f-f_\ell$ and
  $\dot{M}_s^D+\dot{M}_s^R=f_\ell-q_s$. Pressure is a global property
  of the boiler, and therefore, it stands to reason, that it should not
  depend on internal properties of the boiler.\finrema
\end{remark}

\begin{remark}
  On the other hand, unlike pressure, drum water level is not a global
  property of the boiler. It will be shown in the next section that,
  in order to obtain a model for drum water level, it is necessary to
  know other internal quantities such as~$f_s,~f_w$
  and~$f_w^*$.\finrema
\end{remark}

\section{A Model for Drum Water Level}\label{sec:dwlmodel}

Consider a section of the boiler drum as shown in
Fig.~\ref{fig:boiler}, where~$L$ is the steady state (nominal) height
of the water, and~$\delta$ denotes variations around~$L$. Note that,
to determine the height of the water it is necessary to know, not only
the mass of water in the drum,~$M_w^D$, but also the mass of steam
below the water line,~$M_s^{BW}$. An expression for~$M_s^{BW}$ can be
obtained by again applying the conservation of mass principle, i.e.:
\begin{align}\label{eq:MsBW}
\dot{M}_s^{BW}(t) = f_s(t) - f_s(t-a)
\end{align}
where~$f_s(t),~f_s(t-a)$ denote the mass flow of steam out of the
risers, and a delayed version of~$f_s(t)$, respectively. The time
delay~$a$ can be easily computed by noting that it is the time taken
for a given mass to cover a certain distance, i.e.:
\begin{align}
  \label{eq:a}
  a
  =\dfrac{\textrm{distance}}{\textrm{speed}}
  = \dfrac{L+\delta}{f_s(t)/(\rho_sA^R)}
\end{align}
where~$A^R$ denotes the total area of the risers.
\begin{remark}\label{ass:k}
  Equation~\eqref{eq:a} implies that the mass flow of steam below the
  water line follows a straight trajectory from the risers to the
  surface of the drum water. It may prove beneficial to consider a
  free parameter~$k\in\R^+_0$, so that~$f_s(t-k\cdot{}a)$ can account
  for other trajectories. Note that~$k$ would be the only free parameter in
  the model.\finrema
\end{remark}
\begin{assumption}\label{ass:half_full}
  In steady state, the nominal drum water level,~$L$, corresponds to the
  centre of the drum.\finrema
\end{assumption}
An immediate consequence of Assumption~\ref{ass:half_full} is:
\begin{align}\label{eq:DWL1}
  V_w^D + V_s^{BW}
  = \dfrac{V^D}{2} + A^D\delta
\end{align}
where~$A^D$ denotes the area at the centre line of the drum. Using the
fact that~$V=M/\rho$, then equation~\eqref{eq:DWL1} can be rewritten as:
\begin{align}
  \dfrac{M_w^D}{\rho_w} + \dfrac{M_s^{BW}}{\rho_s} = \dfrac{V^D}{2} + A^D\delta\nonumber
\end{align}
Finally, the following expression for the drum water level
deviation,~$\delta$, is obtained:
\begin{align}\label{eq:delta}
 \delta = \left(A^D\right)^{-1} \left( \dfrac{M_w^D}{\rho_w} + \dfrac{M_s^{BW}}{\rho_s} - \dfrac{V^D}{2} \right)
\end{align}

\begin{remark}
  Equations~\eqref{eq:delta},~\eqref{eq:MsBW} and~\eqref{eq:MwD} show
  that it is necessary to be able to independently
  describe~$f_s,~f_w,~f_w^*$ in order to obtain drum water
  level.\finrema
\end{remark}

\begin{remark}
  Note that~\eqref{eq:CVD} and~\eqref{eq:CVR} are linearly dependent
  equations in~$f_s$ and~$f_w$. Therefore, with these two
  equations alone it is not possible to obtain independent expressions
  for~$f_s$ and~$f_w$. This problem will be resolved in
  Section~\ref{sec:newrisers}.\finrema
\end{remark}

\begin{remark}
  An expression for~$f_w^*$ will be derived in
  Section~\ref{sec:momentum} using conservation of momentum 
  in the downcomer-riser system.\finrema
\end{remark}

\section{Spatial Discretisation and Homogeneous Mixing in the
  Risers}\label{sec:newrisers}

In this section, the model of the risers will be embellished to account for
spatial distribution of the boiling process. An additional assumption
will be introduced which allows separation of the expressions for~$f_s$
and~$f_w$ in the model.
\subsection{Spatial discretisation}
Consider a uniform subdivision of the volume of the risers into~$n$
sections. For each section, there are three core equations, describing
mass balance, constant volume and energy balance. This leads to:
\begin{align}
  \dot{M}_s^{R_i} ={}&  f_s^{i-1} - f_s^i +f_\ell^i\label{eq:MsRi}\\
  \dot{M}_w^{R_i} ={}& f_w^{i-1} - f_w^i - f_\ell^i\label{eq:MwRi}\\
    V^{R_i} ={}&  V_w^{R_i} + V_s^{R_i}\label{eq:CVRi}\\
    \dfrac{d}{dt} \left\{     M_s^{R_i}u_s + M_w^{R_i}u_w +  M_m^{R_i}C_pT_m  \right\}
  ={}& (f_s^{i-1}- f_s^i)h_s + (f_w^{i-1}-f_w^i)h_w +Q^{B_i}\label{eq:REBi}
\end{align}
where the superscript~$R_i$ denotes the~$i$-th section of the risers,
~$f_\ell^i$ denotes the mass flow of water converted into steam in
section~$i$, and~$f_s^i,~f_w^i$ denote the mass flow of steam
and water leaving section~$i$, respectively. Similarly,~$f_s^{i-1},~f_w^{i-1}$
denote the mass flow of steam and water entering section~$i$,
respectively.~$Q^{B_i}$ denotes the heat flow directly affecting
section~$i$, for~$i=1,\ldots,n$.

\begin{assumption}
  The heat flow used for boiling water,~$Q^B$, is distributed
  uniformly across the~$n$ sections of the risers,
  i.e.~$Q^{B_i}=Q^B/n,~\forall i$.\finrema
\end{assumption}
The following equations are immediate:
\begin{subequations}
  \label{eq:interface}
\begin{align}
f_w^0={}&f_w^*\\
  f_s^0 ={}&0\\
  f_w^n ={}& f_w \label{eq:fw}\\
  f_s^n ={}& f_s\label{eq:fs}\\
  \sum_{i=1}^n f_\ell^i={}& f_\ell\\
  \sum_{i=1}^n M_w^{R_i}={}& M_w^{R}\\
  \sum_{i=1}^n M_s^{R_i}={}& M_s^{R}
\end{align}
\end{subequations}
To simplify the equations in the sequel, the following
variables are defined:
\begin{align}
  \label{eq:C2i}
  C_2^i \triangleq{}& \dfrac{M_s^{R_i}}{\rho_s^2}\dfrac{\partial\rho_s}{\partial{}P}
                    + \dfrac{M_w^{R_i}}{\rho_w^2}\dfrac{\partial\rho_w}{\partial{}P}\\
    K_2^i \triangleq{}&  M_s^{R_i}\dfrac{\partial{}h_s}{\partial{}P} +
                    M_w^{R_i}\dfrac{\partial{}h_w}{\partial{}P} - V^{R_i} 
 +M_m^{R_i} C_p \dfrac{\partial{}T_s}{\partial{}P}   \label{eq:K2i}
\end{align}
Then, using the same procedure as in Sections~\ref{sec:derivCVR}
and~\ref{sec:derivREB} for equations~\eqref{eq:CVRi}
and~\eqref{eq:REBi}, it follows that:
\begin{align}\label{eq:CVRifinal}
  \dfrac{f_w^i}{\rho_w} + \dfrac{f_s^i}{\rho_s}
  ={}& \dfrac{f_w^{i-1} - f_\ell^i}{\rho_w} + \dfrac{ f_s^{i-1}+f_\ell^i}{\rho_s} 
       - C_2^i\cdot \dot{P}\\
   f_\ell^i  ={}&\left(h_s - h_w\right)^{-1}  \left( - K_2^i\dot{P} + Q^{B_i}
                \right) \label{eq:fellifinal}
\end{align}
Equations~\eqref{eq:CVRifinal} and~\eqref{eq:fellifinal}, together
with~\eqref{eq:MsRi} and~\eqref{eq:MwRi}, give a complete account of
the dynamics of the~$i$-th section of the risers. However,~$f_s^i$
and~$f_w^i$ are still linearly dependent. In the next subsection, an
additional assumption is introduced which allows~$f_s^i$ and~$f_w^i$
to be separately described.

\subsection{Homogeneous mixing in a section of the risers}

Consider a section of the risers. Then, over an infinitesimal period
of time~$\Delta$, the mass of steam and water leaving the section are given
by~$f_s^i\Delta$ and~$f_w^i\Delta$, respectively.
The steam quality of each section is defined as:
\begin{align}
  \label{eq:alphai}
  \alpha^i=\dfrac{M_s^{R_i}}{M_s^{R_i}+M_w^{R_i}}
\end{align}
The following assumption is next introduced:
\begin{assumption}
  Perfect mixing of water and steam occurs in each section of the
  risers.\finrema
\end{assumption}
An immediate consequence of the above assumption is that
the mass of steam and water leaving a specific
section over a period of time~$\Delta$ must have the same
ratio~$\alpha^i$. Therefore,
\begin{align}
    \alpha^i ={}& \dfrac{M_s^{R_i}}{M_s^{R_i}+M_w^{R_i}} =
  \dfrac{f_s^i\Delta}{f_s^i\Delta+f_w^i\Delta} 
\end{align}
Solving for~$f_s^i$ leads to:
\begin{align}\label{eq:fsifinal}
  f_s^i = \dfrac{M_s^{R_i}}{M_w^{R_i}} f_w^i
\end{align}
Substituting~\eqref{eq:fsifinal} into~\eqref{eq:CVRifinal} yields:
\begin{align}
  \dfrac{f_w^i}{\rho_w} + \dfrac{1}{\rho_s} \dfrac{M_s^{R_i}}{M_w^{R_i}} f_w^i
  ={}& \dfrac{f_w^{i-1} - f_\ell^i}{\rho_w} + \dfrac{ f_\ell^i+f_s^{i-1}}{\rho_s} 
       - C_2^i\cdot \dot{P}  
\end{align}
Solving for~$f_w^i$ leads to:
\begin{align}\label{eq:fwifinal}
f_w^i
  ={}&
\left( \dfrac{1}{\rho_w} + \dfrac{1}{\rho_s} \dfrac{M_s^{R_i}}{M_w^{R_i}} \right)^{-1}
\left( \dfrac{f_w^{i-1} - f_\ell^i}{\rho_w} + \dfrac{ f_\ell^i+f_s^{i-1}}{\rho_s} 
       - C_2^i\cdot \dot{P}  \right)
\end{align}

In summary, equations~\eqref{eq:fsifinal} and~\eqref{eq:fwifinal}
provide a separate account of the mass flow of steam and water leaving
the~$i$-th section of the risers. Together with
equations~\eqref{eq:fellifinal},~\eqref{eq:MsRi} and~\eqref{eq:MwRi}
this constitutes a complete description of the dynamics of a section
of the risers.  Using equations~\eqref{eq:interface}, the model for the
sections of the risers can be interfaced with the pressure and drum
water level models presented in Sections~\ref{sec:pmodel}
and~\ref{sec:dwlmodel}.

\begin{remark}
  Note that the concept of homogeneity of the steam-water mix is
  directly related to the concept that no slip occurs between the
  steam mass flow and the water mass flow --
  see~\cite{Astrom2000}. Indeed, the no slip condition implies the
  linear speed of both steam and water masses leaving each section of
  the risers are the same, and therefore, the mass flows must be
  locked together.\finrema
\end{remark}

\section{Water Flow in the Downcomers (Momentum
  Balance)}\label{sec:momentum}
In order to obtain an expression for~$f_w^*$, conservation of momentum
is applied along the downcomers and risers.  The fixed control volume
is defined as the total volume of the downcomer-riser configuration as
shown in Fig.~\ref{fig:boiler}. The control surface is defined as the surface
of the control volume.  A general expression for momentum balance is
given by (see~\cite[Section 2.5]{anderson2017fundamentals}):
\begin{align}
  \label{eq:MB}
  \underbrace{\dfrac{\partial}{\partial{}t} \oiiint_{V} \rho  \bm{\vec{\nu}} dV}_{A}
  + \underbrace{\oiint_S (\rho \bm{\vec{\nu}} \cdot \bm{dS}) \bm{\vec{\nu}}}_{B}
  {}=& \underbrace{-\oiint_S P \bm{dS}}_{C} + \underbrace{\oiiint_V
       \rho{}f dV}_{D} + F_{\mathrm{visc}} 
\end{align}
where the term A denotes the time rate change of the linear momentum of the
contents of the control volume, the term B denotes the nett flow of linear
momentum out of the control surface by mass flow, the term C denotes the
force exerted by pressure on the control surface, the term D denotes the
body force acting on the control volume, and~$F_{\mathrm{visc}}$ denotes the
viscous forces acting on the control surface.

\begin{assumption}
The pressure dynamics are much slower than the momentum
dynamics.\finrema
\end{assumption}
A consequence of the above assumption is that density can be
considered to be uniform across the volume of the downcomers/risers
system.  Therefore,
\begin{align}
  \dfrac{\partial}{\partial{}t} \oiiint_{V} \rho  \bm{\vec{\nu}} dV
  ={}&
    \dfrac{\partial}{\partial{}t} \left\{
  M_w^{DC} \dfrac{f_w^*}{\rho_wA^{DC}} + M_w^R \dfrac{f_w}{\rho_wA^R}  +
  M_s^R \dfrac{f_s}{\rho_sA^R} \right\}\\
  \oiint_S (\rho \bm{\vec{\nu}} \cdot \bm{dS}) \bm{\vec{\nu}}
  ={}&k_{w^*} \dfrac{(f_w^*)^2}{\rho_wA^{DC}}
    + k_w \dfrac{f_w^2}{\rho_wA^{R}}   + k_s \dfrac{f_s^2}{\rho_sA^{R}}\\
  -\oiint_S P \bm{dS}
  ={}&0\\
  \oiiint_V \rho{}f dV
  ={}&  (M_w^{DC} - M_w^R - M_s^R)g\\
  F_{\mathrm{visc}} ={}&0
\end{align}

Note that, since the downcomers contain only water, then
\begin{align}
  \dfrac{M_w^{DC}}{\rho_wA^{DC}} = \dfrac{V_w^{DC}}{A^{DC}} = L^{DC}
\end{align}
where~$L^{DC}$ is the length of the downcomers. Therefore,
equation~\eqref{eq:MB} can be written as:
\begin{align}
      \dfrac{\partial}{\partial{}t} \left\{
  L^{DC} f_w^* + M_w^R \dfrac{f_w}{\rho_wA^R}  +
  M_s^R \dfrac{f_s}{\rho_sA^R} \right\}
  + k_{w^*} \dfrac{(f_w^*)^2}{\rho_wA^{DC}}
  + k_w \dfrac{f_w^2}{\rho_wA^{R}}
  &+ k_s\dfrac{f_s^2}{\rho_sA^{R}}\nonumber\\
  ={}& (M_w^{DC} - M_w^R - M_s^R)g
\end{align}
Solving for~$f_w^*$ leads to:
\begin{align}\label{eq:fw*}
  L^{DC} \dfrac{d}{dt} \left\{f_w^*\right\}
  =
  (M_w^{DC} - M_w^R - M_s^R)g
  - k_{w^*} \dfrac{(f_w^*)^2}{\rho_wA^{DC}}
  - k_w \dfrac{f_w^2}{\rho_wA^{R}}
  - k_s\dfrac{f_s^2}{\rho_sA^{R}}
  \nonumber\\
    -\dfrac{\partial}{\partial{}t} \left\{
   M_w^R \dfrac{f_w}{\rho_wA^R}  +
  M_s^R \dfrac{f_s}{\rho_sA^R} \right\}
\end{align}

\begin{remark}
  Note that~\eqref{eq:fw*} represents a significant departure from the
  equations used in~\cite{Astrom2000}.\finrema
\end{remark}

\section{Superheater Model}
\label{sec:superheater}

A superheater is a heat exchanger used to convert saturated steam
generated in a boiler into superheated steam by adding heat, thus drying
the steam. Superheated steam is used to power turbines to generate
electricity. For the current purpose, the main difference between saturated
and superheated steam is that, when considering saturated steam, it
sufficed to use one state variable, namely the pressure of the
water/steam mixture. This made it possible to unequivocally describe
density, enthalpy, temperature, and other state variables, for both
liquid and vapour phases. However, to describe
the state of superheated steam it is necessary to consider two
independent state variables. In the sequel, pressure and enthalpy will
be used for this purpose.

Let the superheater have volume~$V^{SH}$ and a heat flow
input~$Q^{SH}$. Then  mass balance, energy balance and
constant volume equations for a superheater can immediately be
obtained as shown below.

\subsection{Mass balance}
\begin{align}\label{eq:MsSH}
\dot{M}_s^{SH} = q_s - q_s^{SH} 
\end{align}
where~${M}_s^{SH},~q_s,~q_s^{SH}$ denote the mass of steam in the
superheater, the steam mass flow out of the drum into the
superheater, and the steam mass flow out of the superheater.

\subsection{Energy balance}
\begin{align}  \label{eq:SHEB}
  \dfrac{d}{dt} \left\{M_s^{SH}u_s^{SH}\right\} =  q_sh_s - q_s^{SH}
  h_s^{SH} + Q^{SH}
\end{align}
By definition~$u=h-P/\rho$, therefore,
\begin{align} 
  \dfrac{d}{dt} \left\{M_s^{SH}h_s^{SH} -
  \dfrac{M_s^{SH}}{\rho_s^{SH}} P^{SH}\right\} =  q_sh_s - q_s^{SH}
  h_s^{SH} + Q^{SH} \nonumber
\end{align}
Noting that~${M_s^{SH}}/{\rho_s^{SH}}=V^{SH}$, and expanding the
derivative, leads to:
\begin{align} 
  \dot{M}_s^{SH}h_s^{SH} + M_s^{SH}\dot{h}_s^{SH} - V^{SH}\dot{P}^{SH}
  =  q_sh_s - q_s^{SH}  h_s^{SH} + Q^{SH} \nonumber  
\end{align}
Finally, using equation~\eqref{eq:MsSH} and cancelling the common
terms yields:
\begin{align} \label{eq:SHEBfinal}
   M_s^{SH}\dot{h}_s^{SH} - V^{SH}\dot{P}^{SH}
  =  q_s(h_s-h_s^{SH}) + Q^{SH}
\end{align}

\subsection{Volume Constraint}
\begin{align}\label{eq:CVSH}
\dfrac{d}{dt} \left\{ V^{SH}\right\}  = 0
\end{align}
By definition we know that~$V=M/\rho$, thus
\begin{align}\label{eq:CVSH1}
  \dfrac{d}{dt} \left\{ V^{SH}\right\}
  = \dfrac{d}{dt} \left\{ \dfrac{M_s^{SH}}{\rho_s^{SH}}\right\}
  = \dfrac{\dot{M}_s^{SH}}{\rho_s^{SH}} -
  \dfrac{M_s^{SH}}{(\rho_s^{SH})^2}\dot{\rho}_s^{SH}
\end{align}
However, as mentioned earlier, density of superheated steam is no longer a function of
pressure only. Therefore, the time derivative of density must now be
expanded as follows:
\begin{align}\label{eq:rhodotSH}
  \dot{\rho}_s^{SH} =
  \dfrac{\partial{}\rho_s^{SH}}{\partial{}h_s^{SH}} \dot{h}_s^{SH} +
  \dfrac{\partial{}\rho_s^{SH}}{\partial{}P^{SH}} \dot{P}^{SH}
\end{align}
Substituting~\eqref{eq:CVSH1} and~\eqref{eq:rhodotSH}
into~\eqref{eq:CVSH}, and noting that~$M/\rho=V$ leads to:
\begin{align}
  \dfrac{\dot{M}_s^{SH}}{\rho_s^{SH}} -
  \dfrac{V^{SH}}{\rho_s^{SH}} \left(
  \dfrac{\partial{}\rho_s^{SH}}{\partial{}h_s^{SH}} \dot{h}_s^{SH} +
  \dfrac{\partial{}\rho_s^{SH}}{\partial{}P^{SH}} \dot{P}^{SH}\right) =0\nonumber
\end{align}
Using~\eqref{eq:MsSH} and reordering terms yields:
\begin{align}\label{eq:CVSHfinal}
  {V^{SH}} \left(
  \dfrac{\partial{}\rho_s^{SH}}{\partial{}h_s^{SH}} \dot{h}_s^{SH} +
  \dfrac{\partial{}\rho_s^{SH}}{\partial{}P^{SH}} \dot{P}^{SH}\right)
  = q_s - q_s^{SH} 
\end{align}

\subsection{A model for the superheater}
Equations~\eqref{eq:MsSH},~\eqref{eq:SHEBfinal}
and~\eqref{eq:CVSHfinal} provide a complete model describing the
dynamics of a superheater. To implement the model, equations~\eqref{eq:SHEBfinal}
and~\eqref{eq:CVSHfinal} should be decoupled.
From~\eqref{eq:CVSHfinal}, it follows that:
\begin{align*}
{V^{SH}}\dot{P}^{SH}
  = \left( \dfrac{\partial{}\rho_s^{SH}}{\partial{}P^{SH}}\right)^{-1}
\left(  q_s - q_s^{SH} -    {V^{SH}} 
  \dfrac{\partial{}\rho_s^{SH}}{\partial{}h_s^{SH}} \dot{h}_s^{SH}\right)
\end{align*}
Substituting into~\eqref{eq:SHEBfinal} leads to:
\begin{align*}
  M_s^{SH}\dot{h}_s^{SH} +
   \left( \dfrac{\partial{}\rho_s^{SH}}{\partial{}P^{SH}}\right)^{-1}
  {V^{SH}} 
  \dfrac{\partial{}\rho_s^{SH}}{\partial{}h_s^{SH}} \dot{h}_s^{SH}
  =  q_s(h_s-h_s^{SH}) + Q^{SH} 
  + \left( \dfrac{\partial{}\rho_s^{SH}}{\partial{}P^{SH}}\right)^{-1}
\left(  q_s - q_s^{SH}\right)
\end{align*}
Solving for~$\dot{h}_s^{SH}$ yields:
\begin{align*}
 \dot{h}_s^{SH}
  =
\left(  M_s^{SH} +
   \left( \dfrac{\partial{}\rho_s^{SH}}{\partial{}P^{SH}}\right)^{-1}
  {V^{SH}}  \dfrac{\partial{}\rho_s^{SH}}{\partial{}h_s^{SH}}
  \right)^{-1}
 \left( q_s(h_s-h_s^{SH}) + Q^{SH} 
  + \left( \dfrac{\partial{}\rho_s^{SH}}{\partial{}P^{SH}}\right)^{-1}
  \left(  q_s - q_s^{SH}\right)
  \right)
\end{align*}
Then, from~\eqref{eq:SHEBfinal} it follows that:
\begin{align*}
    \dot{P}^{SH}
  =
  \left( V^{SH} \right)^{-1}
  \left(
  M_s^{SH}\dot{h}_s^{SH} - q_s(h_s-h_s^{SH}) - Q^{SH}
  \right)
\end{align*}
In summary, the model for a superheater is given by the following
equations:
\begin{align}
  \dot{M}_s^{SH} ={}& q_s - q_s^{SH} \\
   \dot{h}_s^{SH}  ={}&
\left(  M_s^{SH} +
   \left( \dfrac{\partial{}\rho_s^{SH}}{\partial{}P^{SH}}\right)^{-1}
  {V^{SH}}  \dfrac{\partial{}\rho_s^{SH}}{\partial{}h_s^{SH}}
  \right)^{-1}
 \left( q_s(h_s-h_s^{SH}) + Q^{SH} 
  + \left( \dfrac{\partial{}\rho_s^{SH}}{\partial{}P^{SH}}\right)^{-1}
  \left(  q_s - q_s^{SH}\right)
  \right) \\
      \dot{P}^{SH}  ={}&
  \left( V^{SH} \right)^{-1}
  \left(
  M_s^{SH}\dot{h}_s^{SH} - q_s(h_s-h_s^{SH}) - Q^{SH}
  \right)
\end{align}

\section{Key consequences of the new model}
\label{sec:simulations}
This section summarises and illustrates the main consequences of the
model derived in this paper. In particular, three key points are made,
namely: (i) drum water level is proportional to steam flow out of the
boiler, (ii) spatial discretisation of the risers is necessary for
fast transient dynamic modelling, and (iii) the relationship between
downcomer mass flow and pressure derivatives leads to a model that can
describe fast transients in the drum water level responses.

In the sequel, the simulations and data presented correspond to Boiler
1 at Proserpine Mill. Boiler~1 has a maximum continuous rating (MCR)
of $17.5[kg/s]$ at~$1650[kPa]$. Boiler~1 does not have a
superheater. The details of the physical parameters used in the
simulation are as follows:
\begin{subequations}
  \label{tab:para}
\begin{align}
A^R={}&1.5~[m^2]\\
V^R={}&10.5~[m^3]\\
A^D={}&9.46~[m^2]\\
V^D={}&10.2~[m^3]\\
L^{DC}={}&7~[m]\\
A^{DC}={}&0.62~[m^2]\\
V^{DC}={}&9.32~[m^3]\\
M_m^D ={}&7400~[kg]\\
M_m^R ={}& 40700~[kg]\\
  P_0={}&1.60\cdot{}10^6~[Pa]\\
  h_w^f={}& 399900~[J/kg]\\
  C_p={}& 470~[J/(kg\cdot{}K)]
\end{align}
\end{subequations}

\begin{remark}
  It is very important to note that the above parameters have all
  been obtained from \emph{physical properties} of the boiler and its
  associated datasheets. No estimation of parameters has
  been performed. This avoids the problem of overfitting due to the
  presence of many degrees of freedom~\cite{ljung2010perspectives}.\finrema
\end{remark}

\subsection{Drum Water Level proportional to Steam Flow}

\begin{figure}[h]
\centering
        \includegraphics[width=\linewidth]{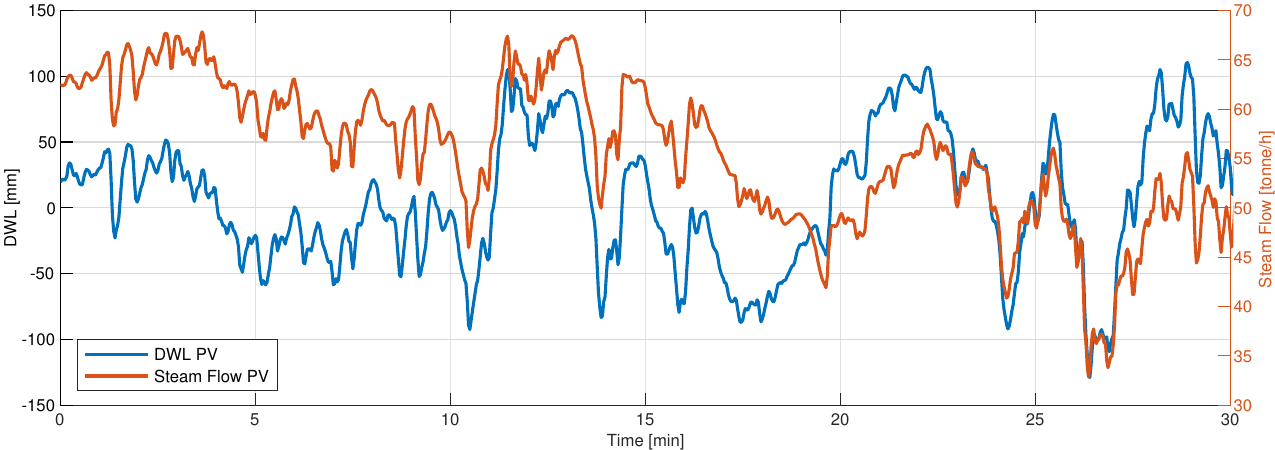}
       \caption{Drum Water Level proportional to Steam Flow}
       \label{fig:prop}
     \end{figure}

One advantage of having a physical model is that particular
occurrences observed in real life can be substantiated by using the
model. Fig~\ref{fig:prop} shows real data from Boiler 1 at Proserpine
Mill for a~$30~[min]$ period. It can be seen that positive changes in
Steam Flow are correlated with positive changes in Drum Water Level
and viceversa. It is hypothesised that this is a general fact that can
be explained by the model. In the following, the model presented in
this paper will be used to show that this hypothesis is, in fact,
true. First it will be proven that the derivative of pressure is
proportional to steam flow, then it will be proven that drum water
level deviations are proportional to the derivative of
pressure. Combining these two observations leads to the final
conclusion that drum water level is indeed proportional to steam flow.

\subsubsection{Derivative of Pressure is proportional to Steam Flow}

Consider equation~\eqref{eq:Pdot}. It can be seen that
equation~\eqref{eq:Pdot} can be rewritten as:
\begin{align}\label{eq:P_lambda_qs}
\dot{P} ={}& -\lambda_1(P,M_s,M_w,V^T)\cdot{}q_s  + \beta(P,M_s,M_w,V^T,q_f,Q^T)
\end{align}
where~$\lambda(\cdot),~\beta(\cdot)$ are nonlinear
functions. Therefore~$\dot{P}$ is proportional to~$q_s$.

\begin{remark}
  Consider the following quantities for Boiler 1 evaluated at the
  nominal operating point:
\begin{align}
  \dfrac{1}{\rho_w} ={}& 1.16\cdot{}10^{-3}\\
  \dfrac{1}{\rho_s} ={}& 1.12\cdot{}10^{-1}\\
  C_3 ={}&  6.16\cdot{}10^{-8}
\end{align}
It can be seen that, in equation~\eqref{eq:Pdot}, the coefficient
multiplying~$q_s$ is at least~$100$ times larger than the others.
This implies that~$q_s$ is the main factor affecting pressure
changes.\finrema
\end{remark}

\subsubsection{Drum Water Level deviations are proportional to
  Derivative of Pressure}
Next, consider equation~\eqref{eq:delta} for the drum water level
deviations~$\delta$ and equation~\eqref{eq:MsBW} for the mass of steam below
the water line~$M_s^{BW}$. It can be seen that~$\delta$ is proportional
to~$M_s^{BW}$. Using Laplace transforms and a Pad\'e approximation for
the time delay, equation~\eqref{eq:MsBW} leads to:
\begin{align*}
  s\cdot{}M_s^{BW}(s)
  ={}& F_s(s)-e^{-as}F_s(s)\\
  ={}& \left(1-\dfrac{2-as}{2+as} \right) F_s(s)\\
  ={}& \dfrac{2as}{2+as} F_s(s)
\end{align*}
where~$s$ is the Laplace Transform variable.  Cancelling the~$s$
(derivative) on both sides of the above equation leads to:
\begin{align}
  M_s^{BW}(s)
  ={}& \dfrac{2a}{2+as} F_s(s)
\end{align}
Using the inverse Laplace transform yields:
\begin{align}
  M_s^{BW}(t)
  ={}&2\cdot{}\int_0^t f_s(\tau) \cdot{}e^{-\frac{2}{a}(t-\tau)}\mu(t-\tau)d\tau
\end{align}
Because of the convolution with the exponential decay, the above
equation can also be written as:
\begin{align}
  M_s^{BW}(t) = 2f_s(t) + \lambda_2(f_s(t-\tau),\tau),\quad 0\leq\tau<{}t
\end{align}
where~$\lambda_2(\cdot)$ denotes the tail of the convolution integral.
Hence, any change in~$f_s(t)$ will appear over a short interval in~$M_s^{BW}(t)$,
i.e. they are proportional.  Finally, consider
equations~\eqref{eq:fsifinal} and~\eqref{eq:fwifinal} for the top
section of the risers, i.e. $i=n$. Then it can be seen that~$f_s(t)$
is proportional to~$f_w(t)$, and that~$f_w(t)$ is proportional
to~$\dot{P}$.

In summary, Drum Water Level deviations are indeed (approximately)
proportional to the Derivative of Pressure.

\subsubsection{Drum Water Level deviations are proportional to Steam Flow}
\label{ssec:prop}
The two facts established in the previous subsections have a major
consequence, namely \emph{Drum Water Level is proportional to Steam
  Flow}. This provides a physical explanation to the experimental
results shown earlier in Fig.~\ref{fig:prop}.

\subsection{Alpha is not a linear function of height in the risers
  under transient conditions}
A common assumption in the literature is that the (mass) steam quality
increases linearly with height in the risers at all times -- see
e.g.~\cite{Astrom2000}. It will be shown
below that, under transient conditions, such an assumption is not valid
and in fact, leads to large errors.

Let $h=7~[m]$ be the height of the risers, and let the risers be
divided in $7$ sections. Let~$\alpha_k,~k=1,\ldots,7$ denote
the (mass) steam quality in each of the~$k$ sections, where~$\alpha_1$
corresponds to the section at the bottom of the risers and~$\alpha_7$
to the section at the top.

Define the ratio~$\bar{\alpha}_k=\alpha_k/(k\cdot\alpha_1)$
for~$k=1,\ldots,7$. If the assumption that~$\alpha_k$ is linear with
height, i.e.~$h/k$, were to be valid, then~$\bar{\alpha}$ would be
equal to~$1,~\forall{}k$, and for all times.

\begin{figure}[h]
  \centering
  \includegraphics[width=\linewidth]{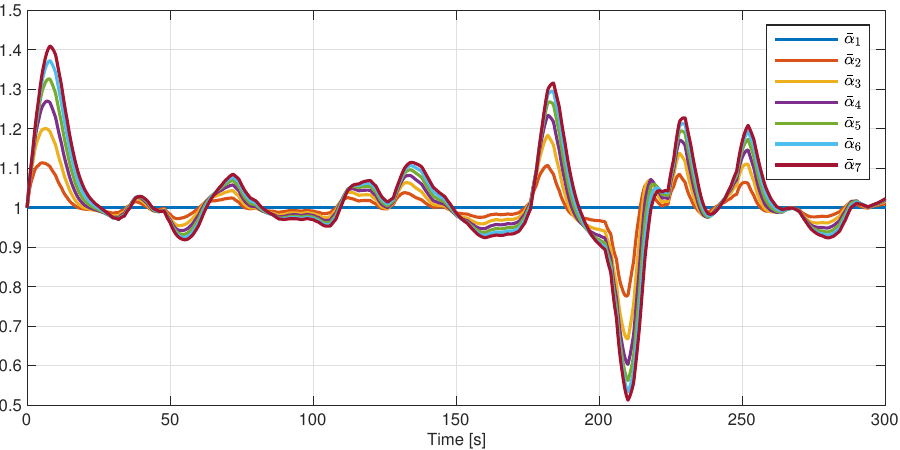}
  \caption{Mass Steam Quality Ratio}
  \label{fig:alpha}
\end{figure}

The full model described in this paper was used to simulate the boiler
response to the real steam flow profile shown in
Fig.~\ref{fig:dataset}. Fig.~\ref{fig:alpha} shows~$\bar{\alpha}_k$
for~$k=1,\ldots,7$, for the first~$300~[s]$. It can be seen
that~$\bar{\alpha}_k=1,~k=1,\ldots,7$ does not hold under transient
conditions, although it does hold in steady state. Under transient
conditions the discrepancy gets larger the further one moves up the
risers. Indeed, at the top of the risers, there is an error of almost
$50\%$ at~$t=210~[s]$ in the maximum steam quality
predicted. Furthermore, the transient response persists for more than
$20~[s]$ after a load change.

This is an important conclusion because the steam quality at the
top of the risers is the main driving factor in the amount of
water and steam entering the drum, and thus, it has a major impact on
drum water level. A significant transient response such as the ones
shown in Fig.~\ref{fig:alpha} cannot be ignored if the goal is to
capture large and fast drum water level excursions.

\subsection{Tracking Fast Changes}
\begin{figure}[h]
  \centering
  \includegraphics[width=\linewidth]{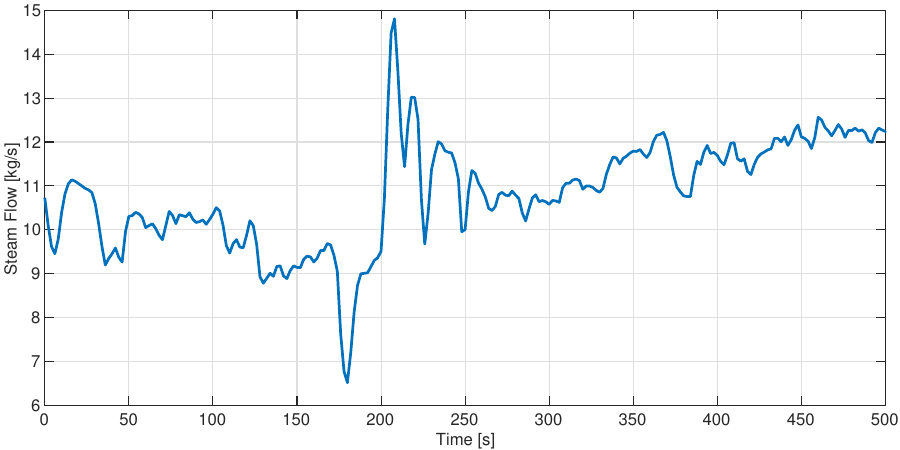}
  \caption{Steam Flow Dataset for simulation}
  \label{fig:dataset}
\end{figure}

The complete model, using the parameters shown in~\eqref{tab:para},
will be used to simulate the Drum Water level response to a steam flow
dataset obtained from Boiler 1 at Proserpine
Mill. Fig.~\ref{fig:dataset} shows this specific steam flow
profile. It can be seen that large steam flow variations occur in a
matter of seconds. In particular, at~$t=200~[s]$ there is a $50\%$
spike in demand which occurs over a period of~$8~[s]$.

\begin{figure}[h]
  \centering
  \includegraphics[width=\linewidth]{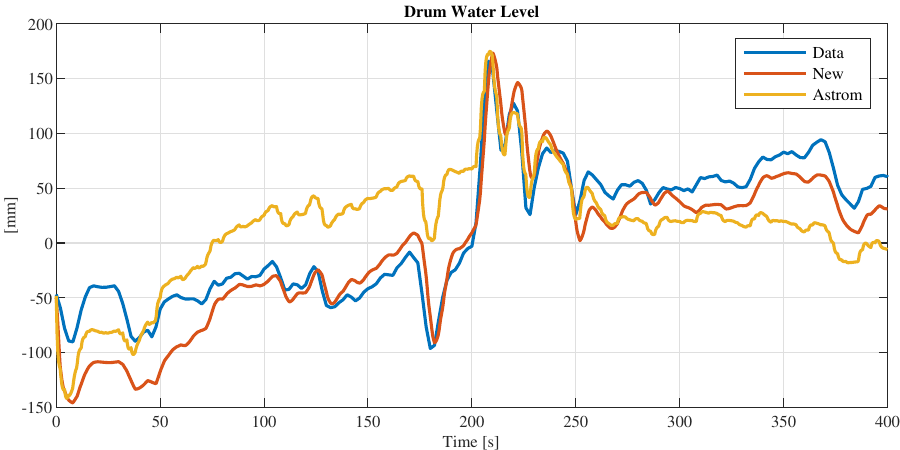}
  \caption{Model Comparison}
  \label{fig:dwlsim}
\end{figure}

As a comparison, the boiler model presented in~\citep{Astrom2000} has
been implemented, fitted and tuned to match Boiler 1 at Proserpine
Mill. The same initial conditions and controllers have been used in
both simulations.  Fig.~\ref{fig:dwlsim} shows a comparison between
the real Drum Water Level response from Boiler 1, and the response
predicted by both models.
It can be seen that the new model accurately tracks the negative peak
at~$t=180[s]$, the positive peak at~$t=208[s]$, and it maintains a
non-increasing offset to the real drum water level data. On the other
hand the model taken from~\cite{Astrom2000} only tracks the positive
peak accurately. As a performance metric, the mean squared error (MSE)
was computed for both models. The MSE for the new model is~$876.4$
whereas the MSE for the \AA{}str\"om and Bell model~\cite{Astrom2000}
is~$2660.8$, i.e. the new model provides an MSE reduction of~$67\%$ in
this particular case.

\section{Implications for Boiler Control}
\label{sec:implications}

This section explores the impact that the new model has on boiler
control architecture and tuning. Two controllers are explored,
namely the steam flow controller and the drum water level controller.

The ideas presented below are based on two key
observations:
\begin{enumerate}[(i)]
\item Drum Water Level deviations are proportional to Steam Flow. Hence,
  if large Steam Flow fluctuations from downstream can be prevented
  from reaching the boiler, the deviations in Drum Water Level can be
  greatly reduced, and
\item Feedwater mass flow cannot be used to correct fast Drum Water
  Level deviations. Considering the geometry of the drum, then the
  maximum available feedwater flow can change the water level at a
  rate of~$2[mm/s]$. The disturbances considered in this paper are of
  the order of~$10[mm/s]$. Thus, the feedwater controller is
  ineffective for fast corrections.
\end{enumerate}
%
Indeed, as mentioned in Section~\ref{sec:intro}, controlling Drum
Water Level under highly variable load conditions has been the main
concern at Proserpine Mill. The insight provided in this section has
proven crucial when developing new steam flow and drum water level
controllers.

\subsection{Steam Flow Controller}

The steam flow controller regulates the opening of the steam valve,
based on an external setpoint and the measurement of the current steam
flow through the valve. The valve is positioned between the drum and
the steam receiver. The mass flow of steam from the boiler to the
steam receiver is proportional to the pressure
difference,~$\Delta{}P$, between them and is also dependent on the
opening of the steam flow valve.

The observation that drum water level deviations are proportional to
steam flow implies that it is highly desirable to prevent large and
rapid steam flow variations from reaching the boiler. Two scenarios
are studied, namely:

\begin{enumerate}
\item When there is a sudden load increase, then pressure in the steam
  receiver will decrease. In turn, this means that~$\Delta{}P$ will
  increase and thus the steam flow out of the boiler will also
  increase. An appropriate control response under these conditions is
  to quickly reduce the opening of the steam flow
  valve. \emph{Therefore the steam flow valve controller time constant
    must be of the same order as the time constant of the steam flow
    perturbations}. The tradeoff associated with this is that there
  will be greater deviations in the steam receiver pressure.
\item When there is a sudden load decrease, then pressure in the steam
  receiver will increase. In turn, this means that~$\Delta{}P$ will
  decrease (possibly to zero) and thus the steam flow out of the
  boiler will also decrease. Unlike the previous scenario, the opening
  of the control valve under this scenario is ineffective as an
  appropriate control response, since no matter how open the valve is,
  the flow of steam is limited by~$\Delta{}P$. Hence another approach
  is needed. One option is to use a let-down valve to release steam
  either to the atmosphere or other independent machinery. Two
  considerations must be made, namely (i) \emph{the letdown valve should
    be located as close as possible either to the source of the steam
    flow perturbation or to the steam receiver}, and (ii) \emph{the
    time constant of the letdown valve controller must be of the same
    order as the time constant of the steam flow perturbations}.
\end{enumerate}
In conjunction, the two scenarios mentioned above provide a viable
strategy for reducing drum water level excursions due to steam flow
variations. The efficacy of these considerations will be illustrated
in Section~\ref{sec:exp} for a boiler in Proserpine Mill.

\subsection{Drum Water Level Controller}
The drum water level controller regulates the feedwater mass flow into
the drum, based on a given setpoint ($0~[mm]$) and the measurement of
the current water level. The largest drum water level perturbation in
Fig.~\ref{fig:dwlsim} has variations of~$70~[mm]$ in~$8~[s]$. This means that
the drum water level can change at a rate of at least $8.75~[mm/s]$.
Due to the geometry of the drum, if $0~[mm]$ is considered to be at
the center of the drum, then an increase of $17~[kg/s]$ in feedwater
(which corresponds to the maximum available flow, from fully closed to
fully open) can only change the Drum Water Level by about $2~[mm/s]$.
In conclusion, fast drum water level disturbances cannot be
compensated with feedwater flow. Specifically, the available control
authority is deficient by a factor of at least 4 : 1.

Traditional drum water level control consists of a classical feedback
controller driven by the error in water level, and a feedforward
controller that uses the measurement of the steam flow out of the
boiler to act ahead of a steam disturbance, i.e. if steam flow
increases then drum water level will increase as well, so the
feedforward will decrease the controller output to preempt the
incoming high water level before slowly increasing the controller
output to balance the new rate at which steam leaves the drum.

\begin{figure}[h]
\centering
        \includegraphics[width=\linewidth]{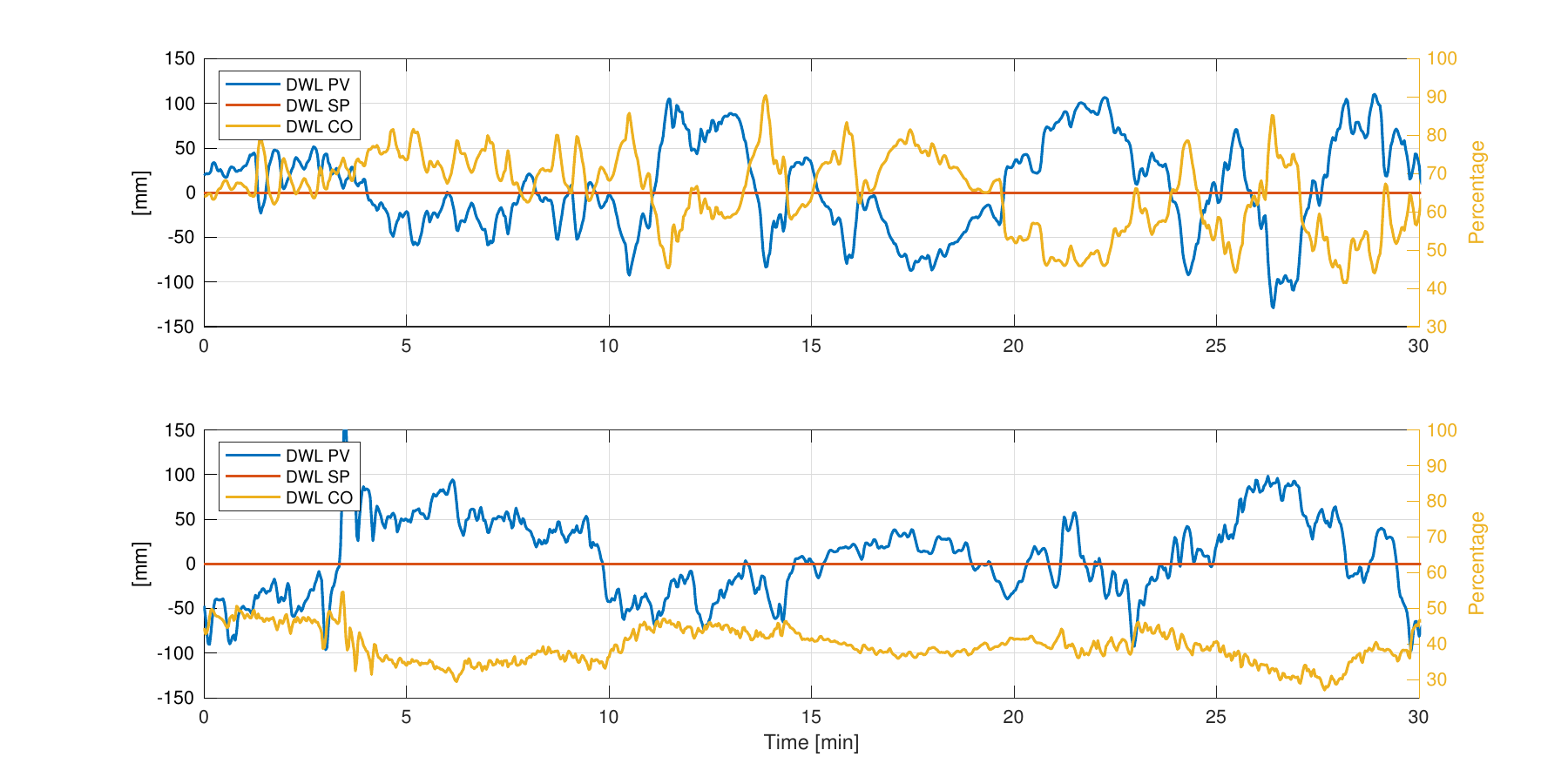}
       \caption{Drum Water Level performance with existing controllers
         (top) and new controllers (bottom)} 
       \label{fig:dwlmirror}
     \end{figure}

The above strategy fails to acknowledge that fast drum water level
excursions can never be compensated with feedwater. Indeed, in
practice, the total controller output is a mirror of the Drum Water
Level measurement, which indicates an almost pure Proportional
controller – see the top plot in Fig.~\ref{fig:dwlmirror}. The problem with this
controller behaviour is that it adds water when the drum water level
is low, but because the reason for it being low is most likely a steam
flow disturbance, it will very likely go in the other direction. When
that happens, the drum water level will be high due to the steam flow
variation, but it will be higher than what it should be because for
the past minute the controller has been putting in more water than
necessary. The same occurs in the other direction. In summary, the
current feedback and feedforward controllers make the Drum Water Level
excursions worse.

The idea behind the new feedforward is that it is impossible to
correct the effects of large load swings on drum water level with
any controller. Therefore, we focus on stopping the feedback
controller from making matters worse, i.e., the new feedforward blinds
the feedback controller to large load swings. The result is that there
is no more mirror behaviour -- see bottom plot in Fig.~\ref{fig:dwlmirror}.

\section{Quantifying Boiler Improvement}
\label{sec:exp}

In order to illustrate the improvements in drum water level deviations
in the day-to-day operation of Boiler 1 at Proserpine Mill six
datasets were compared, each comprising a minimum period of 5
days. Three datasets correspond to boiler operation before the changes
in were made (1-3), and three datasets correspond to after
the fact (4-6). The dates and times in question are as shown in
Table~\ref{tab:dataset}.
\begin{table}[h]
  \centering
  \begin{tabular}[h]{c|ll}
  Dataset & Start time & End time\\
  \hline\hline
1& 07/10/19 at 17:05 & 12/10/19 at 17:05\\
2& 12/10/19 at 17:05 & 17/10/19 at 17:05\\
3& 17/10/19 at 17:05 & 22/10/19 at 17:05\\
4& 16/10/20 at 07:00 & 23/10/20 at 04:00\\
5& 15/07/21 at 21:00 & 21/07/21 at 15:00\\
6& 09/09/21 at 01:00 & 15/09/21 at 21:00    
  \end{tabular}
  \caption{Dataset description}
  \label{tab:dataset}
\end{table}
In the sequel, each dataset will be referenced by their number,
e.g. DS3.
\begin{remark}
  A 5-day dataset includes periods of good and poor boiler operation
  due to fuel moisture changes, and is thus deemed representative of
  normal operation.\finrema
\end{remark}

\begin{figure}[h]
\centering
\includegraphics[width=0.5\linewidth]{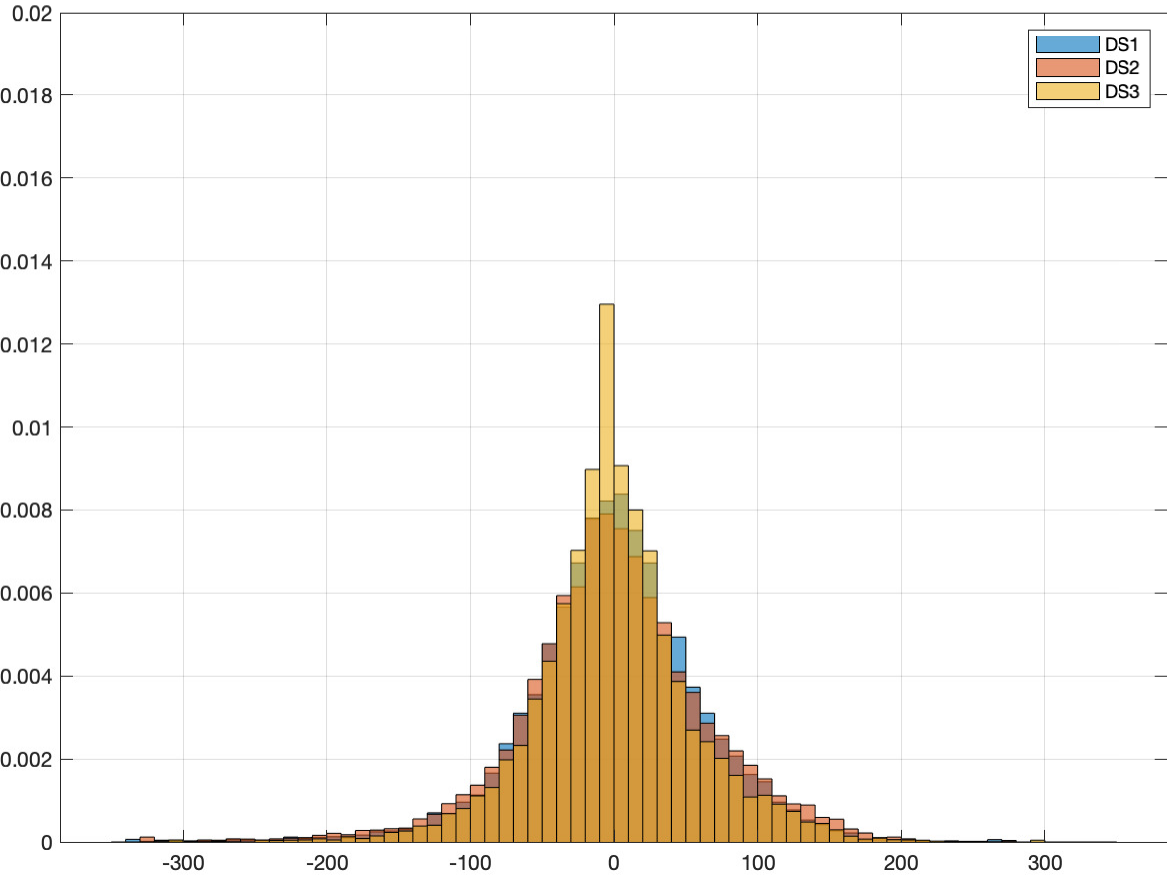}%
        \includegraphics[width=0.5\linewidth]{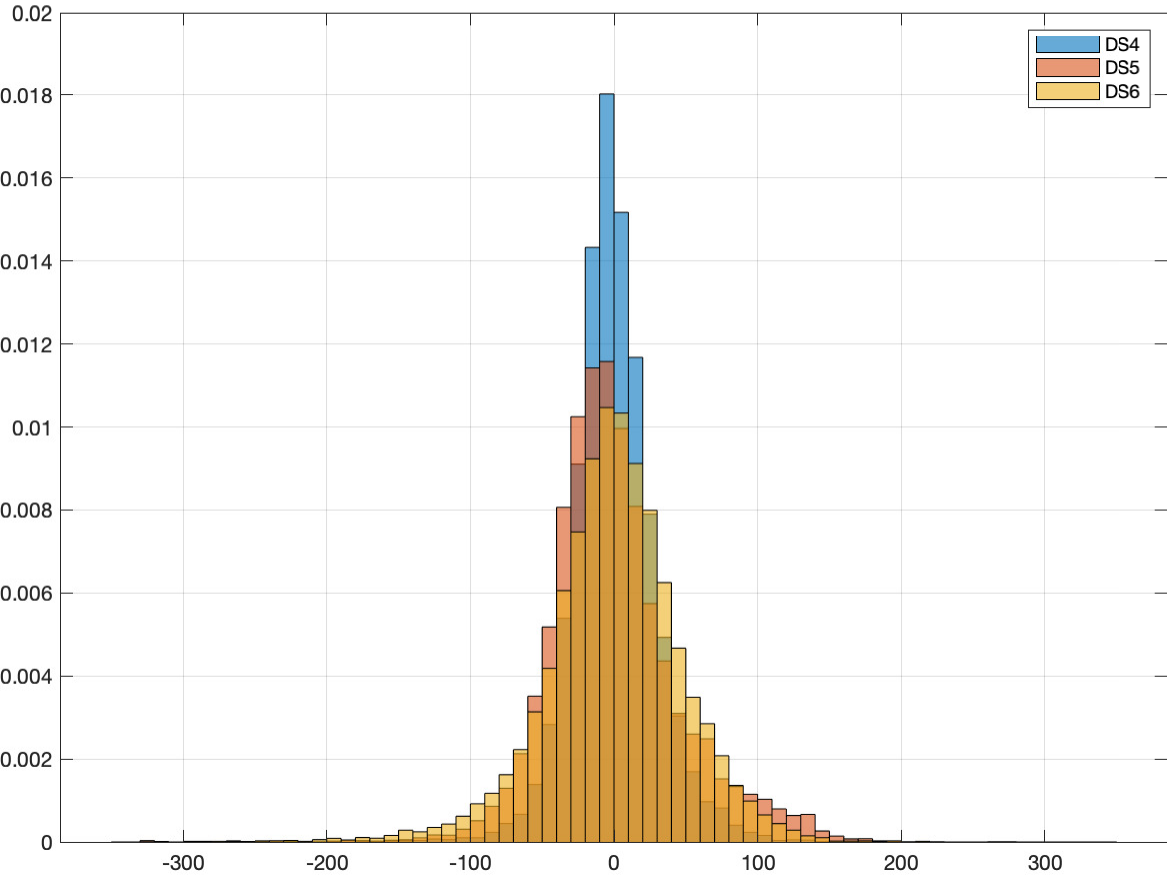}
       \caption{Normalised histogram for Datasets 1-3 (top) and Datasets 4-6 (bottom)}
       \label{fig:histo_norm}
     \end{figure}

Fig~\ref{fig:histo_norm} shows a histogram of Drum Water Level
deviations with mean values removed and where the data in each bin has
been normalised with respect to the total number of datapoints. The
standard deviation of each case is given in Table~\ref{tab:std}.
\begin{table}[h]
  \centering
  \begin{tabular}[h]{cr||cr}
    Dataset & SD&Dataset&SD\\
    \hline\hline
    DS1&60.9 & DS4&28.7\\
    DS2&66.7 & DS5&46.8\\
    DS3&54.9 & DS6&47.6\\
    \hline
    Average & 60.8 & Average & 41.0
  \end{tabular}
  \caption{Dataset standard deviation (DS1-3 existing
    controller. DS4-6 new controller)}
  \label{tab:std}
\end{table}
Several conclusions follow, namely:
\begin{itemize}
\item The histograms presented in Figure~\ref{fig:histo_norm} show
  that the new control laws lead to narrower distribution than those
  corresponding to the original control laws.
\item The above can be interpreted as tighter control over drum water
  excursions resulting from the new controller, i.e. drum water level
  is regulated closer to the setpoint.
\item Table~\ref{tab:std} shows that the new controller has resulted,
  on average, in a~$32.5\%$ reduction in the standard deviation of
  drum water level.
\end{itemize}

Another measure of improvement is to analyse the number of times the
drum water level surpasses a specified safety threshold, e.g.~$\pm
250~[mm]$. Table~\ref{tab:max_excursion} summarises this information.
  \begin{table}[h]
  \centering
  \begin{tabular}[h]{crr||crr}
    Dataset & $<-250$ & $>250$ & Dataset & $<-250$ & $>250$\\
    \hline\hline
    DS1& 27 & 12 & DS4& 0 & 3\\
    DS2& 38 & 15 & DS5& 14 & 2\\
    DS3& 13 & 9 & DS6& 1 & 13\\
    \hline
    Total & 78&36 & Total & 15 &18
  \end{tabular}
  \caption{Water level excursions outside threshold (DS1-3 existing
    controller. DS4-6 new controller)}
  \label{tab:max_excursion}
\end{table}
It can be seen from Table~\ref{tab:max_excursion} that datasets 4-6
(with the new controllers) have a reduction of $80\%$, on average, in
the excursions below~$-250~[mm]$ and a reduction of~$50\%$, on
average, in the excursions above~$250~[mm]$.

\section{Conclusions}
\label{sec:conclusions}

This paper has described a new model for a Boiler operating under
highly variable loads. The model is based on first
principles. Significant departures have been made from the assumptions
previously used in the literature. New features of the model include
(i) a multi-compartment model for the risers, (ii) a new model for
drum water level, and (iii) a new dynamic model for the flow of water
in the downcomers. A comparison between simulations made with the
model and real data from a boiler has been presented which (i) confirm
the validity of the new model, and (ii) highlight the advantages of
the new model under rapid load changing conditions. Implications of
the model for boiler control have also been described with special
emphasis on reducing drum water level excursions under large and rapid
steam flow changes. Experimental results from a boiler at Wilmar
Sugar's Proserpine Mill have confirmed the improvements in drum water
level excursions achieved by the revised control law.

\section*{Acknowledgements}
The authors gratefully acknowledge the extraordinary
help and support from Wilmar Sugar, in particular from Danny
Ferraris, Matt Linneweber, John Andrews and Damien Kelly.

\bibliographystyle{acm}
\bibliography{bibliografia}

\begin{thebibliography}{10}

\bibitem{Alobaid2016}
{\sc Alobaid, F., Mertens, N., Starkloff, R., Lanz, T., Heinze, C., and Epple,
  B.}
\newblock {Progress in dynamic simulation of thermal power plants}.
\newblock {\em Progress in Energy and Combustion Science 59\/} (2016), 79--162.

\bibitem{anderson69}
{\sc Anderson, J.}
\newblock Dynamic control of a power boiler.
\newblock {\em Proceedings of the Institution of Electrical Engineers 116}, 7
  (1969), 1257--1268.

\bibitem{anderson2017fundamentals}
{\sc Anderson, J.}
\newblock {\em Fundamentals of aerodynamics}.
\newblock McGraw-Hill. New York, NY, 2017.

\bibitem{Astrom2000}
{\sc {\AA}str{\"{o}}m, K.~J., and Bell, R.~D.}
\newblock {Drum-boiler dynamics}.
\newblock {\em Automatica 36}, 3 (mar 2000), 363--378.

\bibitem{astrom72}
{\sc {\AA}str{\"o}m, K.~J., and Eklund, K.}
\newblock A simplified non-linear model of a drum boiler-turbine unit.
\newblock {\em International Journal of Control 16}, 1 (1972), 145--169.

\bibitem{Astrom1975}
{\sc {\AA}str{\"{o}}m, K.~J., and Eklund, K.}
\newblock {A simple non-linear drum boiler model}.
\newblock {\em International Journal of Control 22}, 5 (1975), 739--740.

\bibitem{R.D.Bell1987}
{\sc Bell, R.~D., and {\AA}str{\"{o}}m, K.~J.}
\newblock {Dynamic models for boiler-turbine-alternator units data logs and
  parameter estimation for a 160 MW unit}.
\newblock Tech. rep., 1987.

\bibitem{bell+ast96}
{\sc Bell, R.~D., and {\AA}str{\"o}m, K.~J.}
\newblock A fourth order non-linear model for drum-boiler dynamics.
\newblock In {\em IFAC'96, Preprints 13th World Congress of IFAC\/} (San
  Francisco, California, July 1996), vol.~O, pp.~31--36.

\bibitem{dukelow86}
{\sc Dukelow, S.}
\newblock {\em The control of boilers}.
\newblock Instrument Society of America, Research Triangle Park, NC, 1986.

\bibitem{kwatny96}
{\sc Kwatny, K., and Maffezzoni, C.}
\newblock Control of electric power.
\newblock In {\em The Control Handbook}. CRC Press, New York, 1996,
  pp.~1453--1482.

\bibitem{ljung2010perspectives}
{\sc Ljung, L.}
\newblock Perspectives on system identification.
\newblock {\em Annual Reviews in Control 34}, 1 (2010), 1--12.

\bibitem{Maffezzoni1997}
{\sc Maffezzoni, C.}
\newblock {Boiler-turbine dynamics in power-plant control}.
\newblock {\em Control Engineering Practice 5}, 3 (1997), 301--312.

\bibitem{majanne17}
{\sc Majanne, Y., Yli-Fossi, T., Korpela, T., Nurmoranta, M., and Kortela, J.}
\newblock Utilization of drum boilers’ storage capacity for flexible
  operation.
\newblock In {\em 20th IFAC World Congress, Toulouse, France\/} (2017), IFAC.

\bibitem{schulz73}
{\sc Schulz, R.}
\newblock The drum waterlevel in the multivariable control system of a steam
  generator.
\newblock {\em IEEE Transactions on Industrial Electronics and Control
  Instrumentation IECI-20}, 3 (Aug 1973), 164--169.

\bibitem{Sedic2014}
{\sc Sedi{\'c}, A., Katuli{\'c}, S., and Pavkovi{\'c}, D.}
\newblock Dynamic model of a natural water circulation boiler suitable for
  on-line monitoring of fossil/alternative fuel plants.
\newblock {\em Energy conversion and management 87\/} (2014), 1248--1260.

\bibitem{Sunil2017}
{\sc Sunil, P.~U., Barve, J., and Nataraj, P. S.~V.}
\newblock {Mathematical modeling, simulation and validation of a boiler drum:
  Some investigations}.
\newblock {\em Energy 126\/} (2017), 312--325.

\bibitem{Tan2008}
{\sc Tan, W., Fang, F., Tian, L., Fu, C., and Liu, J.}
\newblock {Linear control of a boiler–turbine unit: Analysis and design}.
\newblock {\em ISA Transactions 47}, 2 (apr 2008), 189--197.

\end{thebibliography}

\end{document}